# Altmetrics as traces of the computerization of the research process[1, 2]

Henk F. Moed[3]
Informetric Research Group
Elsevier, Amsterdam, Netherlands
HF.Moed@gmail.com

**To be published, 2016**

**Introduction**

In the Altmetrics Manifesto published on the Web in October 2010, the concept of "Altmetrics" is introduced as follows:

> In growing numbers, scholars are moving their everyday work to the web. Online reference managers Zotero and Mendeley each claim to store over 40 million articles (making them substantially larger than PubMed); as many as a third of scholars are on Twitter, and a growing number tend scholarly blogs. These new forms reflect and transmit scholarly impact: that dog-eared (but uncited) article that used to live on a shelf now lives in Mendeley, CiteULike, or Zotero–where we can see and count it. That hallway conversation about a recent finding has moved to blogs and social networks–now, we can listen in. The local genomics dataset has moved to an online repository–now, we can track it. This diverse group of activities forms a composite trace of impact far richer than any available before. We call the elements of this trace altmetrics. (Priem et al., 2010).

Online reference managers, social networking tools, scholarly blogs, and online repositories are highlighted as technological inventions, and their use by the scientific community or even the wider public leaves traces of impact of scientific activity.

A leading commercial provider of such data, Altmetric.com, distinguishes four types of altmetric data sources (Altmetric.com, 2014):

---

[1] This paper is based on a keynote lecture presented by the author at Altmetrics'14, Indiana University, Bloomington, USA, 23 June 2014.
[2] The contents of this paper express the author's views, and not necessarily those of the author's employer.
[3] Current address: Visiting Senior Fellow, Scuola Superiore degli Studi Avanzati della Sapienza, Sapienza University of Rome, Italy.

- Social media such as Twitter and Facebook, covering social activity;
- Reference managers or reader libraries such as Mendeley or ResearchGate covering scholarly activity;
- Various forms of scholarly blogs reflecting scholarly commentary;
- Mass media coverage, for instance, daily newspapers or news broadcasting services, informing the general public.

I distinguish three drivers of development of the field of altmetrics[4]. *Firstly,* in the policy or political domain, there is an increasing awareness of the multi-dimensionality of research performance, and an increasing emphasis on societal merit, an overview of which can be found in Moed and Halevi (2015a). A typical example of this awareness is the ACUMEN project (Academic Careers Understood through Measurement and Norms) funded by the European Commission, aimed at "studying and proposing alternative and broader ways of measuring the productivity and performance of individual researchers" (Bar-Ilan, 2014). The reader is referred to Bar-Ilan (2014) for an overview of this project and the role of altmetrics therein.

In the domain of technology, a *second* driver is the development of information and communication technologies (ICTs), especially websites and software in order to support and foster social interaction. The technological inventions mentioned in the Altmetrics Manifesto are typical examples of this development. It seems appropriate to link the Altmetrics manifesto to the notion of a "computerization movement". Elliot and Kraemer (2009) define a computerization moment as "… a type of movement that focuses on computer-based systems as the core technologies which their advocates claim will be instruments to bring about a new social order. These advocates of computerization movements spread their message through public discourse in various segments of society such as vendors, media, academics, visionaries, and professional societies" (p. 3). A further positioning of the Altmetrics ideas as computerization movement falls outside the scope of this chapter, even though there is a vast amount of literature on computerization movements, of which Elliot and Kraemer give an overview. I am inclined to conceive the Altmetrics Manifesto as a proclamation of a computerization movement, but a very special one, appealing to basic ideals of science and scholarship. What is important in this chapter is to characterize the type of ideals that inspires the altmetrics movement. I believe they can best be associated with a third driver, primarily emerging from the scientific community itself, namely the Open Science movement. Open Science is conceived as:

> The movement to make scientific research, data and dissemination accessible to all levels of an inquiring society, amateur or professional. It encompasses practices such as publishing open research, campaigning for open access, encouraging scientists to practice open notebook science, and generally making it easier to publish and communicate scientific knowledge. ("Open Science", n.d.).

---

[4] More and more practitioners use the new term "alternative metrics" rather than "altmetrics". In this contribution I use the original term.

The increasing importance of altmetrics is also reflected in the foundation of the NISO Altmetrics Standards Project. The National Information Standards Organization (NISO) is a United States non-profit organization that develops, maintains and publishes technical standards related to publishing, bibliographic and library applications. Funded by the Alfred P. Sloan Foundation, NISO established a project to identify standards and/or best practices related to altmetrics, as an important step towards the development and adoption of new *assessment metrics*. The NISO Project Group published a White Paper in June 2014 (NISO, 2014). [5]

In the NISO Project mentioned above, but also in altmetrics sessions of scientific conferences, altmetrics increasingly linked to—and often limited to—social media references, and to research performance assessment. Empirical studies of altmetrics have focused nearly exclusively on these as well. In Section 2, I will propose a much broader, multi-dimensional conception of altmetrics, namely as *traces of the computerization of the research process*. "Computerization" should be conceived in its broadest sense, including all recent developments in ICT and software, taking place in society as a whole. I distinguish four aspects of the research process: the collection of research data and development of research methods; scientific information processing; communication and organization; and, last but not least, research assessment. I will argue that in each aspect, computerization plays a key role, and metrics are being developed to describe this process. I propose to label the total collection of such metrics as "Altmetrics". In Section 3, I seek to provide a theoretical foundation of altmetrics, based on notions developed by Michael Nielsen in his monograph *Reinventing Discovery: The New Era of Networked Science* (Nielsen, 2011).

---

[5] A project headed by Stefanie Haustein, Cassidy Sugimoto and Vincent Lariviere is being currently funded by the Sloan foundation, to investigate the meanings and motivations of social media metrics. In a way, this activity is a follow-up to the NISO project (Sugimoto, 2014).

To the extent that altmetrics are used as research assessment tools, Section 4 underlines a series of basic theoretical distinctions, which are not only valid in the case of "classical" metrics such as those based on citation analysis, but also, and, perhaps, even more so, in the case of new metrics such as those based on social media references or electronic document usage patterns. These are as follows: the distinction between scientific-scholarly and societal impact; scientific opinion and scientific fact; peer reviewed versus non- peer reviewed manuscripts; immediate and delayed response or impact; intended and unintended consequences of particular behaviors; and, lastly, a distinction between the various domains of science and scholarship, for instance, between natural, technical, formal, biological and medical, social sciences and humanities.

I conclude that altmetrics can provide tools not only to reflect this process passively, but, even more so, to design, monitor, improve, and actively facilitate it. From this perspective, altmetrics can be conceived as tools for the practical realization of the ethos of science and scholarship in a computerized or digital age.

**The Computerization of the Research Process**

I distinguish four aspects of the research process. In this section, I briefly explain these aspects by giving typical outcomes of metrics-based studies of these aspects. The purpose of these examples is to illustrate an aspect, rather than give a detailed account of it. *Firstly,* at the level of the everyday research practice, there is the *collection of research data and the development of research methods*. A "classical" citation analysis in Scopus of articles published during 2002-2012 and cited up until March 2014, generated per discipline a list of the most frequently cited articles. A subject classification of journals was used into 26 research disciplines. It was found that in many disciplines, computing-related articles are the most heavily cited (Halevi, 2014). Table 1 presents nine such articles. The term "computing-related" is used in a broad sense. Most articles describe software packages for data analysis, digital imaging, and simulation techniques. Interestingly, the most frequently cited article in social sciences is about user acceptance of information technology.

**Table 1: Computer science-related top cited articles in Scopus**

| # Cites | Discipline | Article Title |
|---|---|---|
| 17,171 | Agr & Biol Sci, Mol Biol; Medicine | MEGA4: Molecular Evolutionary Genetics Analysis (MEGA) software version 4.0 (2007) |
| 4,335 | Social sciences; business, managemt | User acceptance of information technology: Toward a unified view (2003) |
| 5,325 | Chemistry | UCSF Chimera - A visualization system for exploratory research and analysis (2004) |
| 15,191 | Computer Sci; Eng | Distinctive image features from scale-invariant keypoints (2004) |
| 1,335 | Energy | Geant4 developments and applications (2006) [software for simulating passage of particles through matter] |
| 7,784 | Engineering; Math | A fast and elitist multi-objective genetic algorithm: NSGA-II (2002) |
| 4,026 | Environm Sci | GENALEX 6: Genetic analysis in Excel. Population genetic software….(2006) |
| 4,404 | Materials Science | The SIESTA method for ab initio order-N materials simulation (2002) |
| 10,921 | Physics & Astron | Coot: Model-building tools for molecular graphics (2004) |

The *second* aspect relates to *scientific information processing*. There is a long history of research in the field of information science on information seeking behavior; since this behavior occurs increasingly online, a digital trace of it can be identified. A topic of rapidly increasing importance is the study of searching, browsing, and reading behavior of researchers, based on an analysis of the electronic log files recording the usage of publication archives such as Elsevier's ScienceDirect or an Open Access archive such as arxiv.org. Comparison of citation counts and full text downloads of research articles may provide more insight both into citation practices and in usage behavior (Kurtz et al., 2005; Kurtz & Bollen, 2010; Gorraiz, Gumpenberger, & Schlögl, 2013; Guerrero-Bote & Moya-Anegón, 2014). Table 2 summarizes the main sources of differences between these two types of counts (Moed & Halevi, 2015b). Usage and citation leaks, bulk downloading, differences between reader and author populations in a subject field, the type of document or its content, differences in obsolescence patterns between downloads and citations, and different functions of reading and citing in the research process, all provide possible explanations of differences between download and citation distributions.

**Table 2: Ten important factors differentiating between downloads and citations**

| | | |
|---|---|---|
| 1 | Usage leak: Not all downloads may be recorded. |
| 2 | Citation leak: Not all citations may be recorded. |
| 3 | Downloading the full text of a document does not mean that it is read. |
| 4 | The user (reader) and the author (citer) population may not coincide. |
| 5 | Distribution # downloads less skewed than that of # cites, and depends upon the type of document differently |
| 6 | Downloads and citations show different obsolescence functions. |
| 7 | Downloads and citations measure distinct concepts. |
| 8 | Downloads and citations may influence one another in multiple ways. |
| 9 | Download counts are more sensitive to manipulation. |
| 10 | Citations are public, usage is private. |

*Communication and organization* is a *third* group of aspects. These two elements are distinct, from an altmetric point of view, to the extent that the first takes place via blogs, Twitter and similar social media, whereas the second occurs for instance in scholarly tools as Mendely or Zotero. In this paper, the two aspects will be discussed jointly. The analysis of the use of online tools such as social media, reference managers and scientific blogs perhaps constitutes the core of studies of the computerization in this domain. Many altmetric studies cover this aspect. In a recent special altmetrics issue of the journal *Research Trends*, Thelwall gives an historical overview of the study of social web services using altmetrics, focusing on Mendeley and Twitter (Thelwall, 2014). He underlines the need to further validate altmetrics, by investigating the degree at which they correlate with—or predict—citation counts and other traditional measures. In the same issue, Shema presents an additional, and state of the art, altmetric data source: scholarly blogs (Shema, 2014). The studies focusing on this aspect aim to deepen our understanding of the ways in which researchers communicate and organize themselves, and how the new technologies not only influence communication and organization, but also how they could improve these processes.

The use of altmetrics—or metrics in general—in *research assessment* is a *fourth* aspect of the computerization of the research process. Mentions of authors and their publications in social media like twitter, in scholarly blogs and in reference managers form the basis of the exploration of new impact measures. In his historical overview, Thelwall concludes that "altmetrics [also] have the potential to be used for impact indicators for individual researchers based upon their web presences, although this information should not be used as a primary source of impact information since the extent to which academics possess or exploit social web profiles is variable" and that, "more widely, however, altmetrics should not be used to help evaluate academics for anything important because of the ease with which they can be manipulated" (Thelwall, 2014).

Moed and Halevi (2015a) underline that indicators that are appropriate in one context may be invalid or useless in another. The decision as to which indicators should be used in a particular assessment depends upon a) what units have to be assessed; b) which aspect of research performance is being assessed; c) what constitutes the overall objective of the assessment. The authors introduce the notion of a "meta-analysis" of the units under assessment, in which metrics are not used as tools to evaluate individual units, but rather to reach policy inferences regarding the objectives and general set-up of an assessment process. For instance, publication counts and average journal impact factors of a group's publications are hardly useful in a relative assessment of research active groups with a strong participation in international networks, but they may be very useful in a context in which there is solid evidence that a substantial number of groups is hardly research active or publishing mainly in national journals (Moed & Halevi, 2015a).

**A Theoretical Foundation: Michael Nielsen's "Reinventing Discovery"**
Fully capturing the notion of the ethos of science and scholarship and tracing back its history requires a full essay, the presentation of which reaches far beyond the scope of the current chapter and also exceeds the competency of its author. Perhaps it is appropriate to refer to Francis Bacon and his proposal "for an universal reform of knowledge into scientific methodology and the improvement of mankind's state using the scientific method" ("Francis Bacon", n.d.).

It must be noted that Bacon is generally conceived of as the founder of the positive, empirical sciences. But the ethos I seek to capture does not merely relate to this type of science, but to science and scholarship in general, including, for instance, hermeneutic scholarship. In any case, Bacon's proposal develops *two* base notions, namely the notion that science can be used to improves the state of mankind, and that it is governed by a strict scientific-scholarly methodology. Both dimensions, the practical and the theoretical-methodological, are essential in his idea.

A key issue nowadays is how the ethos of science and scholarship, admittedly outlined so vaguely above, must be realized in the modern, computerized, or digital age. The state of development of information and communication technology (ICTs) creates enormous possibilities for the organization of the research process, as well as for society as a whole. I believe that it is against this background that the emergence and potential of altmetrics should be considered.

Michael Nielsen's (2010) monograph presents a systematic, creative exploration of the actual and potential value of the new ICT for the organization of the research process. The aim of the remaining part of this section is to summarize some of the main features of this thinking. I believe it provides an adequate framework in which altmetrics can be positioned and further developed, without claiming that alternative frameworks are of no value.

In building up his ideas, Nielsen borrows concepts from several disciplines, and uses them as building blocks or models. A central thesis is that online tools can and should be used in science to amplify collective intelligence. Collective intelligence results from an appropriate organization of collaborative projects. In order to further explain this, he uses the concept of 'diversity', borrowed perhaps from biology, or its sub-branch, ecology, but in the sense of cognitive diversity, as he states: "To amplify cognitive intelligence, we should scale up collaborations, increasing cognitive diversity and the range of available expertise as much as possible" (Nielsen, 2010, p. 32).

As each participant can give only a limited amount of attention in a collaboration, there are inherent limits to size of the contributions that participants can make. At this point the genuine challenge of the new online tools comes into the picture: they should create an "architecture of attention," and in my view one of the most intriguing notions in Nielsen's work, "that directs each participant's attention where it is best suited—i.e., where they have maximal competitive advantage."(Nielsen, 2010, p. 33).

In the ideal case, scientific collaboration will achieve what he terms as "designed serendipity," so that a problem posed by someone who cannot solve it finds its way to one with the right micro expertise. Using a concept stemming from statistical physics, namely, critical mass, he further explains that "conversational critical mass is achieved and the collaboration becomes self-stimulating, with new ideas constantly being explored" (Nielsen,2010, p.33).

One of the ways to optimize the collaboration is by modularizing it. Here Nielsen adopts the open source software development as a model. Actually, he speaks of open source collaboration, in which participants work in a modular way, make small contributions, and have easy reuse of earlier work. And, last but not least, this type of collaboration uses signaling mechanisms (e.g., scores, or metrics) to help people to decide where to direct attention.

Also, he uses the concept of "data web," being defined as "a linked web of data that connects all parts of knowledge," and "an online network intended to be read by machines." He underlines that data driven intelligence is controlled by human intelligence and amplifies collective intelligence. Nielsen highlights the potential of the new online tools to stimulate interaction and even collaboration between professional researchers and the wider public, and the role this public can play for instance in data collection processes using crowdsourcing techniques.

My proposal is to use Michael Nielsen's set of creative ideas as a framework in which altmetrics can be positioned. Their role would not merely be that of rather passively descriptors, but, actively, or proactively, as tools to establish and optimize Nielsen's "architecture of attention", a configuration that combines the efforts of researchers and technicians on the one hand, and the wider public and the policy domain on the other. I will further discuss this issue in Section 5. In the next section I will highlight a series of distinctions that are crucial when discussing the potential and limits of altmetrics in the assessment of research performance.

**Useful Distinctions**

To further explore the potential and limitations of altmetrics, I believe it is useful to highlight a series of distinctions that are often made in the context of the use of "classical" metrics and publishing, but that are in my view most relevant in connection with altmetrics as well.
First of all, a most relevant distinction is that between *scientific-scholarly and societal merit* and impact. These two aspects do not coincide. In Section 3, speaking of the ethos of science, two dimensions were highlighted: a practical and a theoretical-methodological: science potentially improves the state of mankind, and is governed by strict scientific-scholarly methodology. I defend the position that these methodological rules are essential to the scientific method. These rules are constitutive for science and scholarship, and discriminate between what is a justified scientific-scholarly knowledge claim and what is not.

Societal merit of scientific–scholarly research is in my view a legitimate and valuable aspect, not only in connection with motives and strivings of individual researchers, but also related to funding and assessment criteria. But it cannot be assessed in a politically neutral manner. To be successful, the project proposed by Bacon and so many others requires a certain distance and independence from the political domain, and most of all, a strong, continuous defense of proper methodological rules when making knowledge claims and examining their validity.

A next distinction is perhaps even more difficult to make, namely between *scientific opinion and scientific fact* or result. In journal publishing, many journals distinguish between research articles on the one hand, and opinion pieces, discussion papers, or editorials on the other. At least in the empirical sciences, the first type ideally reports on the outcomes of empirical research conducted along valid methodological lines, and discusses their theoretical implications. The second type is more informal, normally not peer-reviewed, and speculative. The two types have from an epistemological point of view a different status. I believe it is crucial to keep this in mind when exploring the role of altmetric data sources containing scholarly commentaries, such scientific-scholarly blogs.

At this point, it is also important to distinguish between speculations or opinion pieces related to scientific-scholarly issues, and those primarily connected with political issues. I believe that it is in the interest of the ethos of science to be especially alert to a practice in which researchers make political statements using their authority as scientific-scholarly experts. Such practices should be rigorously unmasked whenever they are detected.

*Intended versus unintended consequences* of particular behavior is a next distinction. During the past ten years or so, the general debate on the application of "classical" metrics based on publication and citations, especially their large-scale use in national research assessment exercises, strongly focused on the effects that the actual use of such metrics have upon researchers, and on the degree of manipulability of the metrics. These were among the main topics of the discussions on the organization of national research assessment exercises in the UK and in Australia. The least that can be said is that this debate is equally relevant as regards the use of altmetrics based on social media. But, as indicated in Section 2, Thelwall warns that the problem of manipulability is much larger in case of altmetrics than it is in the application of citation indices (Thelwall, 2014).

Finally, it is also crucial to distinguish the *various domains of science and scholarship*, for instance, natural, technical, formal, biological, medical, social sciences, and humanities. Although such subject classifications suffer from a certain degree of arbitrariness, it is important to realize that the research process, including communication practices, reference practices, and orientation towards social media, may differ significantly between one discipline and another. In this context one of the limitations of the model Michael Nielsen proposes in his monograph *Reinventing Discovery* should be highlighted: the use of the open source software development as a model of collaboration may fit the domain of the formal sciences rather well, but may be less appropriate in many subject fields in humanities and social sciences. In other passages in his monograph he is aware that this organizational model may not be appropriate in all domains of science and scholarship.

**Concluding Remarks**

What then are the main conclusions of this chapter? I propose a broad conception of altmetrics. Altmetrics is more than measuring attention in social media to scientific-scholarly artifacts, but should be conceived as metrics of the computerization of the research process in general. I propose the set of ideas developed by Michael Nielsen as a framework within which altmetrics can be positioned and further explored. His work represents a thorough, systematic account of the potential of online tools in the research process, and, in this way, articulates the practical realization of the ethos of science and scholarship in the computerized or digital age. He shows how the new online tools support open science, the notion that is in my view one of the pillars, perhaps even the most important one, of the altmetrics manifesto.

Many proponents of altmetrics may, either as a first impression, or after reflection, not be so happy with my proposal. After all, the demarcation between altmetrics and "classical" metrics is rather vague. Citation indexes are also the product of the ICT development, be it in an earlier phase than the current one. Moreover, citation indices are even used to illustrate the computerization of the research process. Therefore, in a sense, classical metrics are altmetrics as well. Both classical metrics and altmetrics are subjected to the same danger, namely, that their utility is limited to a few very specific cases, and both types of metrics do have in principle the same potential.

In the same way that classical citation metrics are often uniquely linked to the use of journal impact factors for assessing individual researchers—although so many other citation-based metrics and methodologies have been developed, applied to different aggregations and with different purposes—altmetrics runs perhaps a danger of being too closely linked with the notion of assessing individuals by counting mentions in Twitter and related social media, a practice that may provide a richer impression of impact than citation counts do, but that has clearly its limitations as well (e.g., Cronin, 2014).

Altmetrics and science metrics, or indicators in general, are much more than that. Apart from the fact that much more sophisticated indicators are available than journal impact factors or Twitter counts, these indicators do not have a function merely in the evaluation of research performance of individuals and groups, but also in the study of the research process. In this way, in terms of a distinction developed in Geisler (2000), these indicators are used as process indicators rather than outcome measures. Also, like science metrics in general, altmetrics does not merely provide reflections of the computerization of the research process, but can, in fact, develop into a set of tools tool to further shape, facilitate, design, and conduct this process.

**Cited References**